\documentclass{elsart}

\usepackage{amssymb}
\usepackage{graphicx}
\usepackage{epsfig}

\begin{document}

\begin{frontmatter}



\hfill {\bf TTP04-27}

\title{Charge asymmetry and radiative  $\phi$ decays
 \thanksref{labelf}}
 \thanks[labelf]{Work supported in part by 
 EC 5-th Framework EURIDICE network  project HPRN-CT2002-00311,
 TARI project RII3-CT-2004-506078 and BMBF/WTZ/POL01/103. }


\author[label1]{Henryk Czy\.z,}
\author[label1]{ Agnieszka Grzeli{\'n}ska}
\author[label2]{ and Johann H. K\"uhn}

\address[label1]{Institute of Physics, University of Silesia,
PL-40007 Katowice, Poland.}
\address[label2]{Institut f\"ur Theoretische Teilchenphysik,
Universit\"at Karlsruhe, D-76128 Karlsruhe, Germany.}

\begin{abstract}
The study of radiative $\phi$ decays into scalar mesons, with subsequent
 decay into $\pi\pi\gamma$, constitutes an important topic at the
 electron-positron collider DAPHNE. The interference of the respective 
 amplitude with the one for $\pi^+\pi^-\gamma$ production, where
 the photon originates from initial state radiation, will allow for unambiguous
 tests of models for the $\phi\to\gamma f_0(\to \pi\pi)$ amplitude.
 The forward-backward asymmetry of charged pions, which is a clear signal of
 such an interference, amounts up to 30 \% and is at the same time
 quite sensitive to the details of the various amplitudes.
 The results for several characteristic cases are presented and their
 implementation into the Monte Carlo generator PHOKHARA is described.

\end{abstract}

\end{frontmatter}

\section{Introduction}
\label{intro}

The method of the radiative return, i.e. the study of events with
 initial state radiation (ISR) at electron-positron colliders,
 allows to measure the hadronic cross section over a wide 
range of energies in a single experiment, without changing the beam energy.
 Originally suggested already
long time ago \cite{Zerwas}, the idea has been revived in \cite{Binner:1999bt},
after high luminosity electron-positron colliders, specifically $\phi$-
 and $B$-meson factories, came into operation. Already now the method
 has been used to study the pion form factor at DAPHNE \cite{KLOE2}
 and the production of $J/\psi(\to \mu^+\mu^-)$ \cite{Babar1} and of 
 $\pi^+\pi^-\pi^0$ at BaBar \cite{Babar2}. It relies on the factorization
 into a part describing the production of the virtual photon from
 the initial state (including radiative corrections) and a part 
 describing the production of hadrons from the virtual photon,
 which essentially corresponds to the cross section for 
 $e^+e^-\to{\mathrm {hadrons}}$. This important feature is affected
 by contributions from the emission of photons from the final state (FSR).
 As discussed in \cite{Binner:1999bt}, suitable kinematic cuts can be chosen,
 which strongly reduce these effects in
 $e^+e^-\to\pi^+\pi^-\gamma$. For $B$-meson factories, configurations
 with hadrons of low invariant mass in one hemisphere and a hard photon
 in the opposite hemisphere are completely dominated by ISR, 
 with a small admixture from the ``two step'' process 
 $e^+e^-\to\gamma\gamma^*(\to \pi^+\pi^- \gamma )$.
 The situation is more involved at the $\phi$-factory DAPHNE. For photons
 at small angles with respect to the beam direction and centrally
 produced pions, ISR is again dominant. This is the kinematic
 region studied in \cite{KLOE2} for a measurement of the pion form factor.
 In general, however, FSR will be more important at these low energies.
 Unavoidably, the input for any FSR analysis and Monte Carlo simulation
 will be model dependent and a careful experimental test of the model
 assumptions is required. The most recent and detailed analysis
 of this problem \cite{Czyz:PH03} has been performed with the Monte Carlo 
 event generator PHOKHARA \cite{PHOKHARA}, where FSR for pions is modeled
 by scalar QED (sQED) (a different ansatz, leading to similar conclusions
 is presented in \cite{Giulia}). This ansatz is certainly adequate for
 most of the kinematic regions and for cms energies away from the $\phi$ 
 resonance. For $\sqrt{s}\simeq 1.02$ GeV, on top of the  $\phi$-resonance,
 an additional complication arises: the presence of the narrow 
  $\phi$-resonance leads to an enhancement of $\pi\pi\gamma$ final states
 through radiative $\phi$-decays into (dominantly scalar) resonances.
 This process is important in itself and has, for many years, been the
 subject of theoretical investigations  (see 
 \cite{BCG,Lucio,Achasov,Achasov1,Mel,Pennington,Maiani}
  and references therein), where the decay
 chain $\phi\to (f_0(980)+f_0(600))\gamma \to \pi \pi \gamma$
 is generally assumed to dominate. First investigations, where this
 channel has been linked to the radiative return, have been published 
 in \cite{Mel}. Two distinct features allow to discriminate the 
 radiative $\phi$ decay from $\pi\pi\gamma$ production through the
 radiative return on one hand and from FSR from charged pions 
 on the other hand: Radiative $\phi$ decays to $\gamma f_0$ will
 lead to $\pi^+\pi^-\gamma$ and $\pi^0\pi^0\gamma$ states with
 a relative weight 2:1; this is in contrast to FSR from pions, which,
 similarly to ISR, leads by construction to $\pi^+\pi^-\gamma$ only.
 FSR is, furthermore, strongly enhanced for photons collinear 
 to the pions. A second important feature, which will be the main subject
 of this paper, is the charge asymmetry. It arises from the interference
 between ISR with $\pi^+\pi^-$ in an odd charge conjugate state, and
 FSR and  radiative  $\phi$ decays, with $\pi^+\pi^-$ in an even 
 charge conjugate state. With the ISR amplitude being already fairly
 well determined, this interference will allow to pin down size
 and even the relative phase of the $\phi\to \pi\pi\gamma$ amplitude
 in a model independent way. As a third possibility the dependence of
 all this distributions on the cms energy and the energy dependent relative 
 importance of $\phi$ and continuum amplitude could be used. However,
 we will not dwell further on this last, fairly obvious, aspect.

\section{Description of the model}
\label{sec1}

 The subsequent analysis will not attempt to study the most general
 amplitude for the radiative $\phi$ decay. Instead we will use
 characteristic models with different resonance admixture and variable phases
 and demonstrate that these will lead to marked differences in the charge 
 asymmetry.

 For definiteness, we have adopted two models describing
 $\phi\to \pi \pi \gamma$ decays ($\pi$ stands for both neutral 
 and charged modes), used also in \cite{Mel}:
 the ``no structure model'' of \cite{BCG} and the $K^+K^-$ model
 of \cite{Lucio}. The models were extended to include both
 $f_0(980)$ and $f_0(600)$ ($\sigma$) intermediate states
 in the decay chain $\phi\to (f_0(980)+f_0(600))\gamma \to \pi \pi \gamma$,
 an extension required to fit experiments 
  \cite{KLOE1,SND}. More sophisticated
 version of the $K^+K^-$ model, incorporating also 
 the complete $f_0(980)+f_0(600)$
 mixing matrix, can be found in \cite{Achasov,Achasov1}.
 Contributions from the $\phi\to \pi \rho(\to \pi\gamma)$ decay chain
 are not taken into account. Indeed KLOE data  \cite{KLOE1} show, that they are
 negligible for the $\pi^0\pi^0\gamma$ final state. 
 The branching ratios BR($\rho^\pm\to\pi^\pm\gamma$)
 and BR($\rho^0\to\pi^0\gamma$) are comparable
 \cite{PDG},
 hence the same applies to the  $\pi^+\pi^-\gamma$
 final state. 

 At present this choice of  models
 is not based on definite experimental information, which is still quite poor,
 but on their simplicity.
 To discriminate
 between different models, more experimental work is needed.
   However, as shown in the next section,
 a lot can be learned from studies of charge and charge induced asymmetries.

 The amplitude for FSR based on sQED, denoted ${\mathcal M}_\mathrm{sQED}$,
 is identical to the one used in PHOKHARA and is given by
 \begin{eqnarray}
 {\mathcal M}_\mathrm{sQED} &=&
 \frac{i e^3}{s}  F_{2\pi}(s) \ 
  \bar v(p_1)\gamma_\mu u(p_2) \
\biggl\{\left(q_1+k-q_2\right)^{\mu} 
 \frac{q_1\cdot\epsilon^*(\gamma)}{q_1\cdot k}\nonumber \\
 &&+\left(q_2+k-q_1\right)^{\mu} 
 \frac{q_2\cdot\epsilon^*(\gamma)}{q_2\cdot k}
 -2\epsilon^{*\mu}(\gamma)\biggr\} \ .
\label{sQED}
\end{eqnarray}

 The NLO correction to FSR and the complete description of ISR,
 including NLO corrections, is also identical to the one used
 in PHOKHARA and described in \cite{PHOKHARA,Czyz:PH03,Radcor}.
 The pion form factor $F_{2\pi}$ is taken from \cite{Khodja}.

 For the sake of definiteness, we write down explicitely  the amplitude
 $\mathcal M_\phi$
  for $\phi\to (f_0(980)+f_0(600))\gamma \to \pi \pi \gamma$ 

 \begin{eqnarray}
 {\mathcal M}_\phi = -i e f_\phi(Q^2) \epsilon_\mu(\phi)d^{\mu\alpha}
 \epsilon^*_\alpha(\gamma), \ \ \ \ 
  d^{\mu\alpha}= (P.k)g^{\mu\alpha}-k^\mu P^\alpha \ ,
\label{mphi}
\end{eqnarray}

\noindent
which, when combined with $\phi$ production, describes the
 resonant process 
 $e^+e^- \to  \phi^*\to (f_0(980)+f_0(600))\gamma \to \pi \pi \gamma$
 with the amplitude

\begin{eqnarray}
 \kern-15pt
 {\mathcal M}_{e^+e^-}(\phi) = \frac{i e^3}{s}\bar v(p_1)\gamma_\mu u(p_2)
  \  e^{i\alpha_\phi} \ 
 \frac{g_{\phi\gamma} }{M_\phi^2-s-iM_\phi\Gamma_\phi} \
 \ f_\phi(Q^2) \ d^{\mu\alpha}\epsilon^*_\alpha(\gamma) \ .
\label{mee}
\end{eqnarray}

\noindent
 Here $P$ ($\epsilon(\phi)$) and $k$ ($\epsilon(\gamma)$) are $\phi$ and 
 photon four momenta (polarization vectors) respectively. $P=p_1+p_2=Q+k$
 in the case of $e^+(p_1)e^-(p_2)$ annihilation amplitudes, $q_1$ ($q_2$) is
 the $\pi^+$ ($\pi^-$) four momentum and  $Q=q_1+q_2$.
 $M_\phi$ denotes $\phi$ mass and $\Gamma_\phi$ its width.
 Unless stated otherwise, all numerical values of physical parameters
  are taken from \cite{PDG}.
 The $g_{\phi\gamma}$ coupling, deduced from $\Gamma(\phi\to e^+ e^-)$,
 is $g_{\phi\gamma} = 7.765 \cdot 10^{-2}$ GeV$^2$.
 The phase $\alpha_\phi$
 in $M_{e^+e^-}(\phi)$ becomes relevant for the interference between
$M_{e^+e^-}(\phi)$ and both ${\mathcal M}_\mathrm{sQED}$ and ISR
 amplitudes
 and thus can be measured for charged pions.
 It is kept constant in the `no structure' model. In the case of the 
 $K^+K^-$ model it is taken from
 \cite{Achasov1}: $\alpha_\phi = b \sqrt{s-4m_\pi^2}$, with 
 $b=75^\circ /  GeV$
   (thus also constant for fixed $s$).
 
 The function
 $f_\phi(Q^2)$ reads

\begin{eqnarray}
 f_\phi(Q^2) = \frac{g_{\phi f_0 \gamma}g_{f_0\pi\pi}}
 {M_{f_0}^2-Q^2-iM_{f_0}\Gamma_{f_0}} +
 e^{i\alpha_\sigma} \frac{g_{\phi \sigma \gamma}g_{\sigma\pi\pi}}
 {M_{\sigma}^2-Q^2-iM_{\sigma}\Gamma_{\sigma}} \ ,
\label{fphins}
\end{eqnarray}

\noindent
in the `no structure' model, and

\begin{eqnarray}
 f_\phi(Q^2) &=& \frac{g_{\phi K^+K^-}}{2\pi^2 m_K^2} \ \ 
 I\left(\frac{M_\phi^2}{m_K^2},\frac{Q^2}{m_K^2}\right)\nonumber \\
 &&\cdot \left[\frac{ g_{f_0 K^+K^-}g_{f_0\pi\pi}}
 {M_{f_0}^2-Q^2-iM_{f_0}\Gamma_{f_0}} +
 e^{i\alpha_\sigma} \frac{ g_{\sigma K^+K^-}g_{\sigma\pi\pi}}
 {M_{\sigma}^2-Q^2-iM_{\sigma}\Gamma_{\sigma}}\right] \ ,
\label{fphikk}
\end{eqnarray}

\noindent
in the $K^+K^-$ model. The meaning of the three-particle couplings
 $g_{m_1m_2m_3}$
is self-ex\-plan\-a\-to\-ry and, assuming isospin symmetry, these couplings 
 are  identical for charged and neutral pion modes.
 $M_{f_0}$ and $\Gamma_{f_0}$ ($M_\sigma$ and $\Gamma_{\sigma}$)
  are the mass and the width of $f_0(980)$ ($f_0(600)$).
 Again for simplicity, we use constant widths for both resonances.  
 The phase $\alpha_\sigma$ has to be substantially different from zero
 to fit the experimental data \cite{KLOE1,SND} (see below). $I(\cdot,\cdot)$
 is the K-loop function given in \cite{Lucio}.

\begin{table}
\label{t:tab1}
\begin{center}
\caption{The values of the parameters obtained by fits
 to the experimental data \cite{KLOE1,SND}}
\begin{tabular}{ccc}
  & `no structure' model  & \ \ \ `$K^+K^-$' model \\ \hline 
 $M_{f_0}$ [MeV] & $983.9\pm 1.4$  & $1003\pm 8$ \\
 $\Gamma_{f_0}$ [MeV]& $35.4\pm 3.2$  & $108\pm 7$ \\ 
 $M_\sigma$ [MeV]&$588\pm 145$ & $519\pm 23$\\
 $\Gamma_{\sigma}$  [MeV] & $1653\pm 921$ &$319\pm 65$ \\
 $\alpha_\sigma$ & $(101\pm 11)^\circ$& $(98\pm 8)^\circ$ \\
 $g_{\phi f_0 \gamma}g_{f_0\pi\pi}$ & $2.84 \pm 0.17$&  -\\ 
 $g_{\phi \sigma \gamma}g_{\sigma\pi\pi}$&$4.0 \pm 1.5$ & - \\
 $ g_{\phi K^+K^-} g_{f_0 K^+K^-}g_{f_0\pi\pi}$ [GeV$^2$]&- & $59.4 \pm 5.1$\\
 $ g_{\phi K^+K^-} g_{\sigma K^+K^-}g_{\sigma\pi\pi}$ [GeV$^2$]&- &
 $7.9 \pm 2.0$ \\ \hline
\end{tabular}
\end{center}
\end{table}

 From the amplitude ${\mathcal M}_\phi$ (eq. \ref{mphi}) we have calculated
 the differential 
  branching ratio 
 $dBr(\phi \to\pi^0\pi^0\gamma)/dM_{\pi\pi}$, as measured in \cite{KLOE1,SND}
 ($\sqrt{Q^2}\equiv M_{\pi\pi}$).
 The parameters obtained in the fits
 to these data are shown in Table 1. 
  The values of masses and widths obtained in the fits do depend
 on the choice of the model and at this point it is difficult to interpret
 them as physical masses and widths of the resonances.

 \begin{figure}[ht]
 \vspace{0.5 cm}
\includegraphics[width=12.5cm,height=8cm]{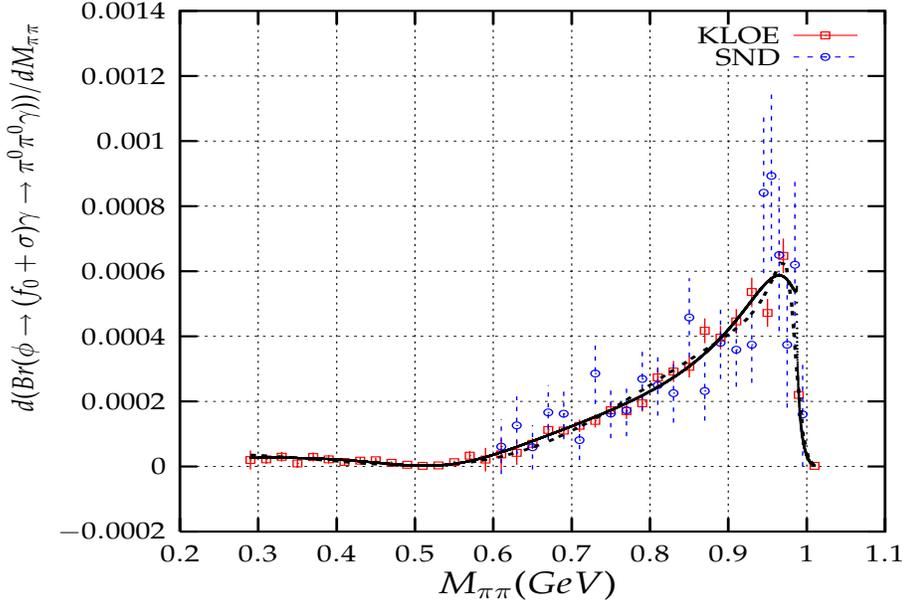}
\caption{Fits to the measured \cite{KLOE1,SND} differential
 branching rations: $K^+K^-$ model - solid line
 ($\chi^2 = 29/(53 \ d.o.f.)$), `no structure' model -
 dashed line ($\chi^2 = 38/(53 \ d.o.f.)$).
}
 \vspace{0.5 cm}
\label{f1}
\end{figure}

 Both models agree well with the data, as shown in Fig.\ref{f1},
 and the destructive interference between $f_0(600)$ and
  $f_0(980)$ amplitudes is needed in both cases
 to describe the low $M_{\pi\pi}$ tail of the distribution.
  As expected, the $\phi\to\pi^0\pi^0\gamma$
 data alone cannot differentiate between different models. However, as
 shown in the next section, charge asymmetries have an enormous
 analyzing power, and allow to test
  details of the radiative $\phi$
 decays to charged pions.

\section{ The analyzing power of charge asymmetries}
\label{sec2}

 The importance of charge asymmetries for tests of the FSR model
  was discussed in details
  in the context of scalar QED in \cite{Binner:1999bt,Czyz:PH03}.
 At the $\phi$-factory DAPHNE, the direct radiative $\phi$-decays,
 not included in \cite{Binner:1999bt,Czyz:PH03},
  play, however an important role \cite{Mel}. These a~priori
 small effects are enhanced by the resonant behavior of $\phi$ and
  their accurate analysis is compulsory. The interest in these asymmetries
 is twofold. On one hand they play an important role in the FSR
 'background' estimate for the measurement of
  the pion form factor 
 via the radiative return method. On the other hand, as we will show below,
 they allow for powerful tests of models describing radiative $\phi$-decays.
 The asymmetry analysis is thus a source of rich experimental information
 complementary to the cross section measurement.
 \begin{figure}[ht]
 \vspace{0.5 cm}
\includegraphics[width=12.5cm,height=9cm]{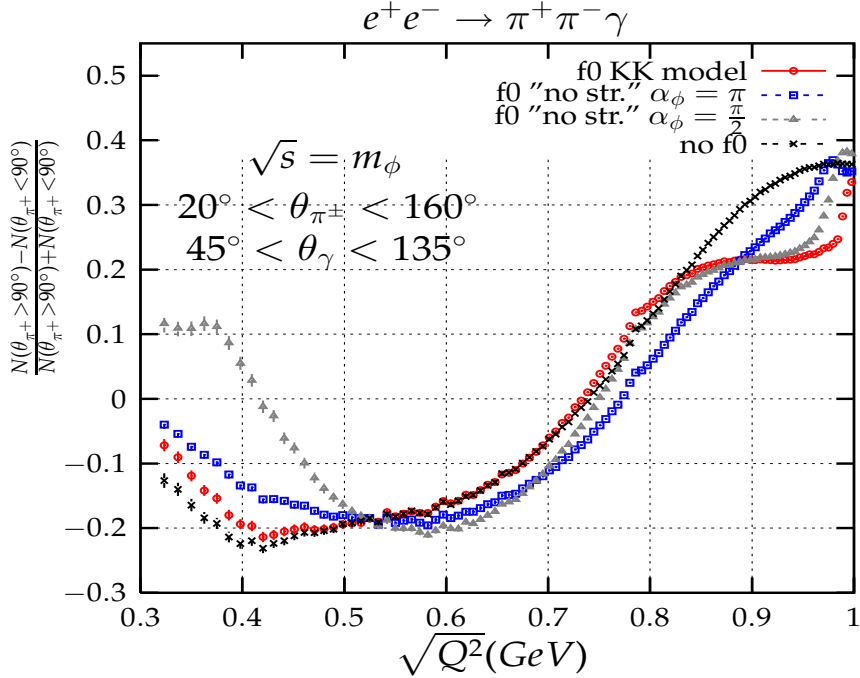}
\caption{Forward-backward asymmetry for different radiative $\phi$
 decay models compared with the asymmetry calculated within sQED (no $f_0$)
}
 \vspace{0.5 cm}
\label{f2}
\end{figure}

 \begin{figure}[ht]
\includegraphics[width=7cm,height=7cm]{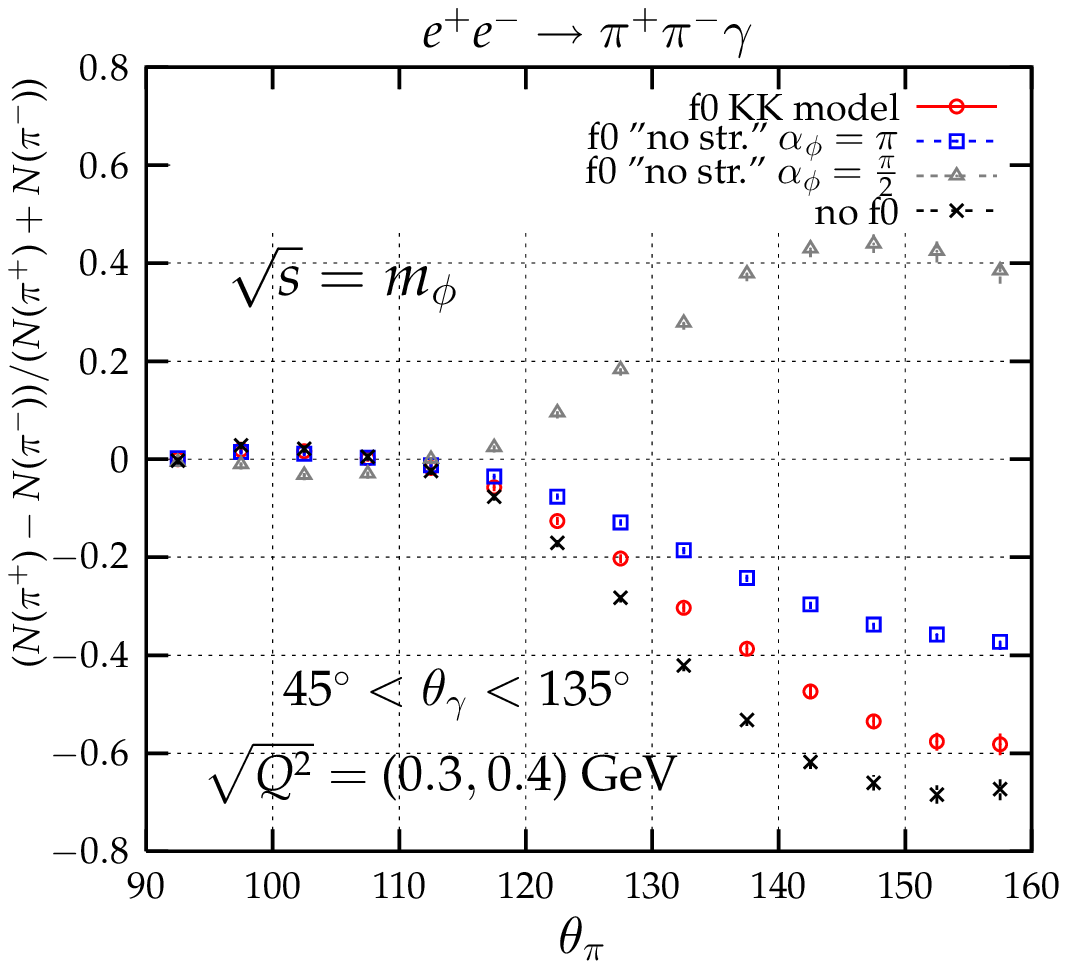}
\includegraphics[width=7cm,height=7cm]{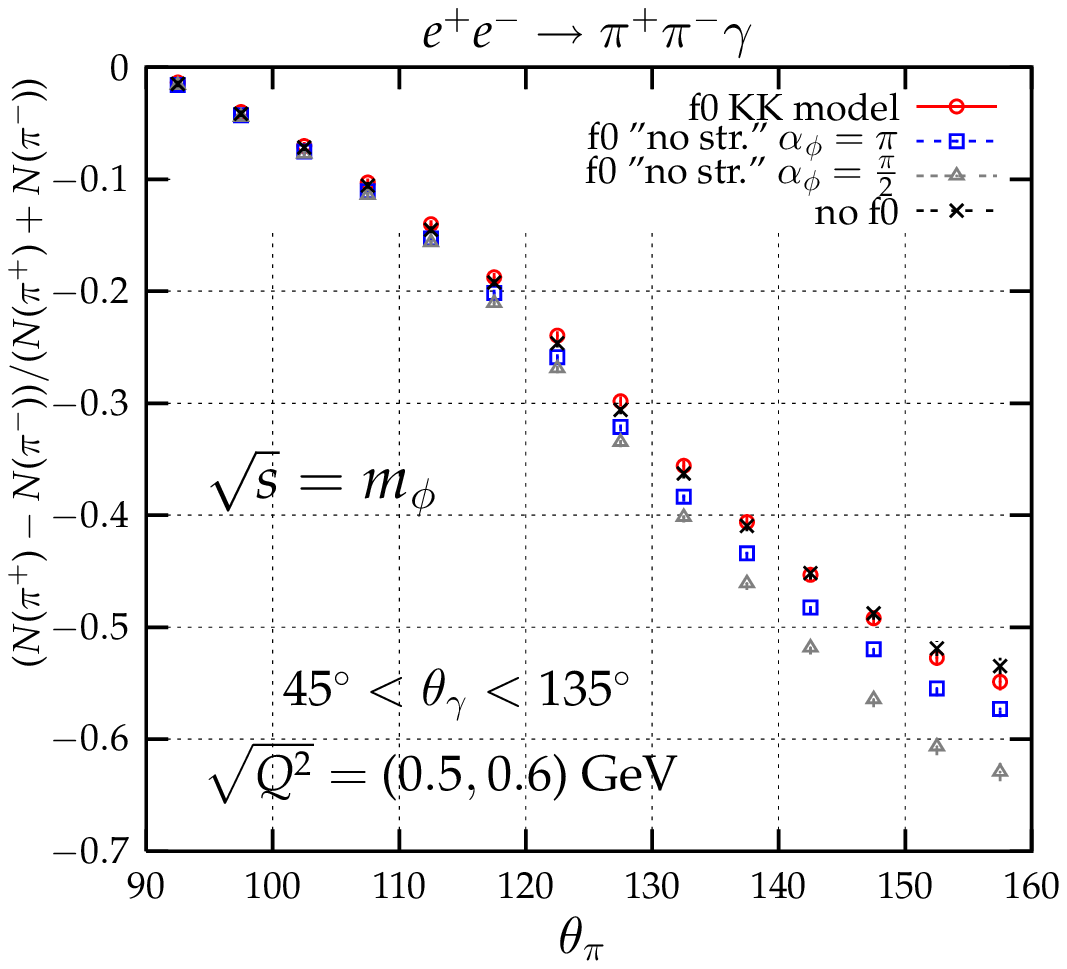}\\
\includegraphics[width=7cm,height=7cm]{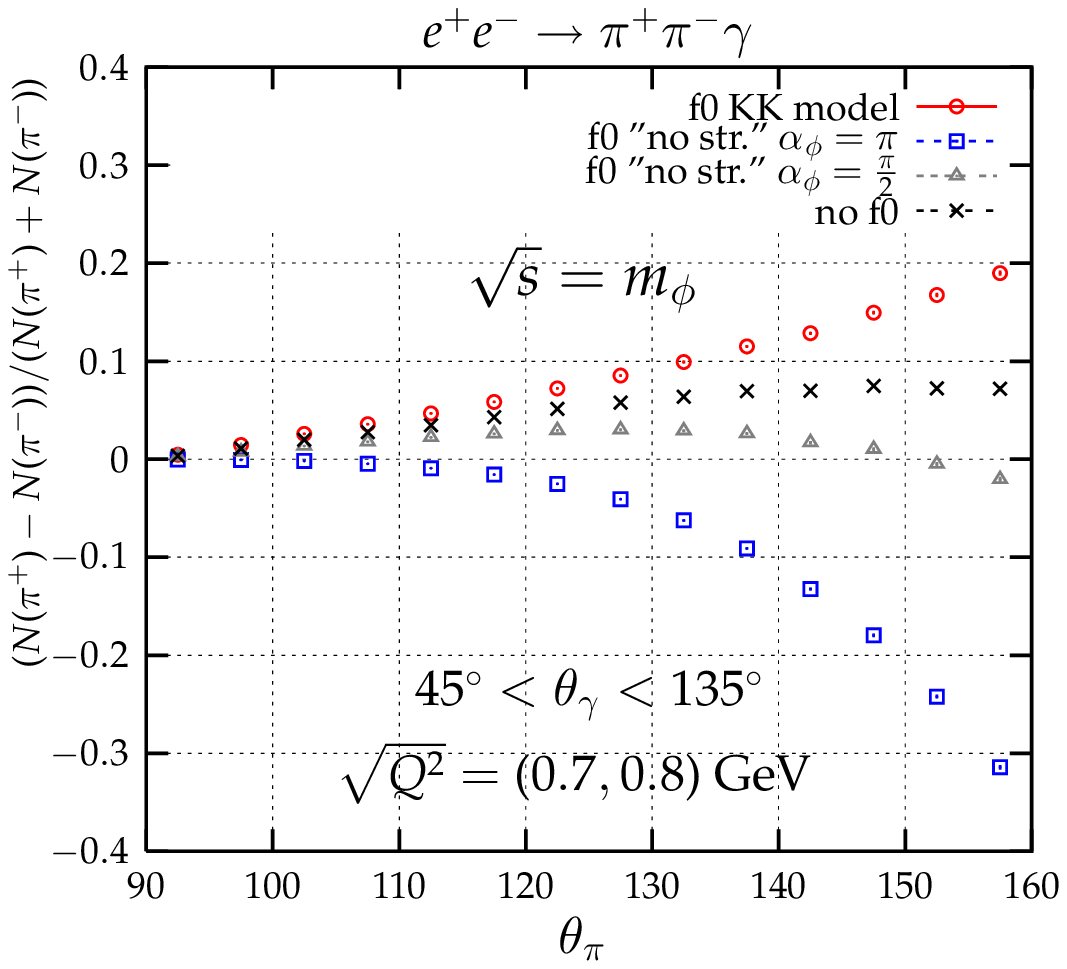}
\includegraphics[width=7cm,height=7cm]{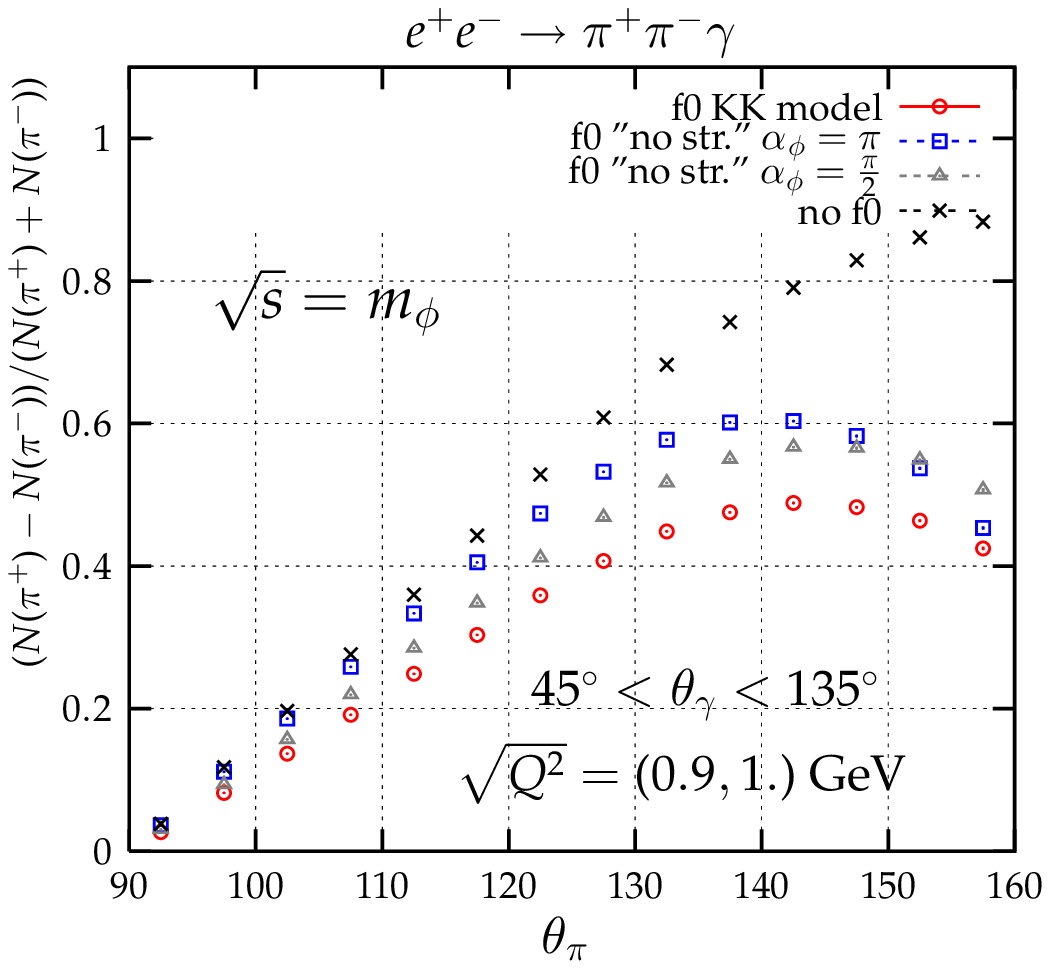}
\caption{Angular and $Q^2$ dependence of charge asymmetries 
for different radiative $\phi$
 decay models compared with the asymmetry calculated within sQED (no $f_0$),
 for centrally produced photons and pions}
\vskip 1cm
\label{f3}
\end{figure}

Let us start with the forward-backward asymmetry defined for $\pi^+$

\begin{eqnarray}
 {\mathcal A}_{FB}(Q^2) = 
 \frac{N(\theta_{\pi^+}>90^\circ)- N(\theta_{\pi^+}<90^\circ) }
 {N(\theta_{\pi^+}>90^\circ)+ N(\theta_{\pi^+}<90^\circ)}\left(Q^2\right) \ ,
\label{asym}
\end{eqnarray}
\noindent
 In this analysis the z- axis is chosen along initial positron direction.
 The cuts on the pion angles $20^\circ < \theta_{\pi^\pm} < 160^\circ$
 are required for pions to be observed in the detector
  and the photon angular range 
 $45^\circ<\theta_\gamma< 135^\circ$ was chosen to enhance the FSR effects.
  The results are shown in Fig. \ref{f2}.
 The asymmetry is very sensitive to the relative phase
 between sQED and direct $\phi$ decay amplitudes ($\alpha_\phi$) 
 and predictions
 of this phase \cite{Achasov1} can be definitely tested.
 The assumption of a  $Q^2$ 
 dependent phase from the complex loop function 
  $I\left(\frac{M_\phi^2}{m_K^2},\frac{Q^2}{m_K^2}\right)$ 
 can also be tested, since the  interference pattern 
 can be measured as a function of $Q^2$.
 The effects depend also on the photon angular range and more detailed studies
 are possible.  Sizable effects are predicted
 and, as one can qualitatively see from preliminary KLOE results \cite{Micco}
 on
 asymmetries, some of the models we use may already be excluded by the data.

 The angular and the $Q^2$ dependence of the charge asymmetry
 for different models are shown in Fig. \ref{f3}.
 Again large effects are observed, with marked differences between
 different models.  The combination of 
 cross section measurements for both charged and neutral pion final
 states
  with  asymmetry measurements
 in the charged mode will allow for the most comprehensive
 study of the $\phi \to \pi\pi\gamma$ decay. 
 \begin{figure}[ht]
 \vspace{0.5 cm}
\includegraphics[width=7cm,height=7cm]{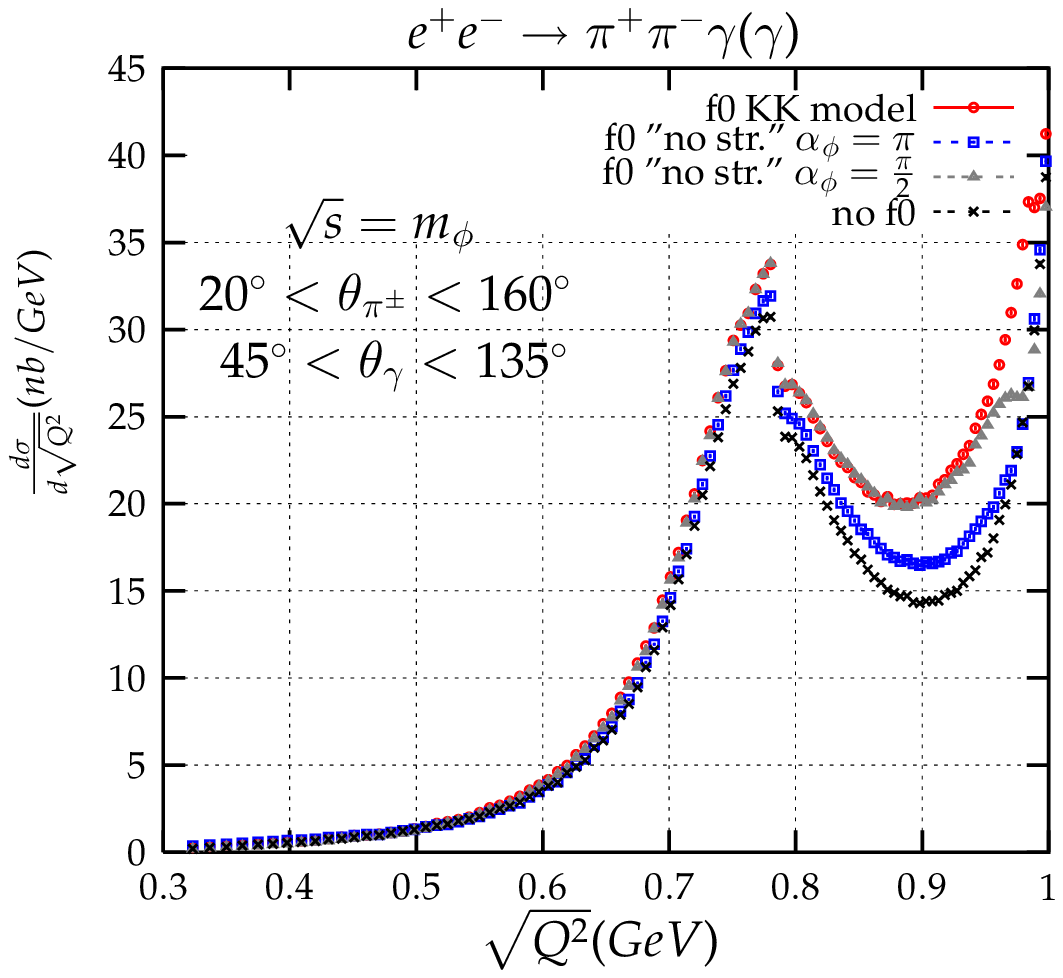}
\includegraphics[width=7cm,height=7cm]{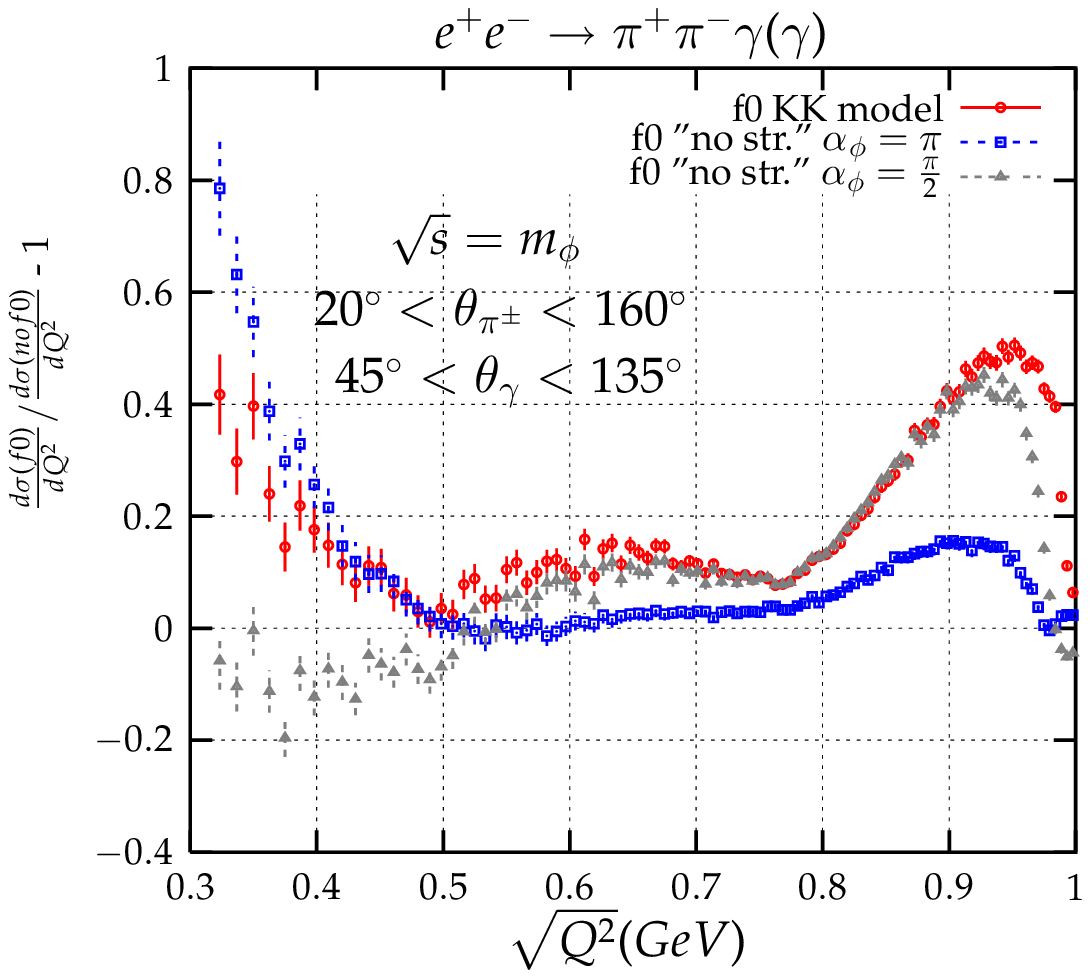}
\caption{Differential cross section and for different radiative $\phi$
 decay models compared with sQED (no $f_0$) cross section (left plot)
 and their relative difference (right plot).}
 \vspace{0.5 cm}
\label{f4}
\end{figure}

The effects of the direct radiative $\phi$ decay on the differential cross
 section are shown in Fig. \ref{f4} for centrally
 produced photons and pions. They are big and the question
 arise if they can affect the KLOE \cite{KLOE2} analysis of the pion
 form factor.
 \begin{figure}[ht]
 \vspace{0.5 cm}
\includegraphics[width=12.5cm,height=9cm]{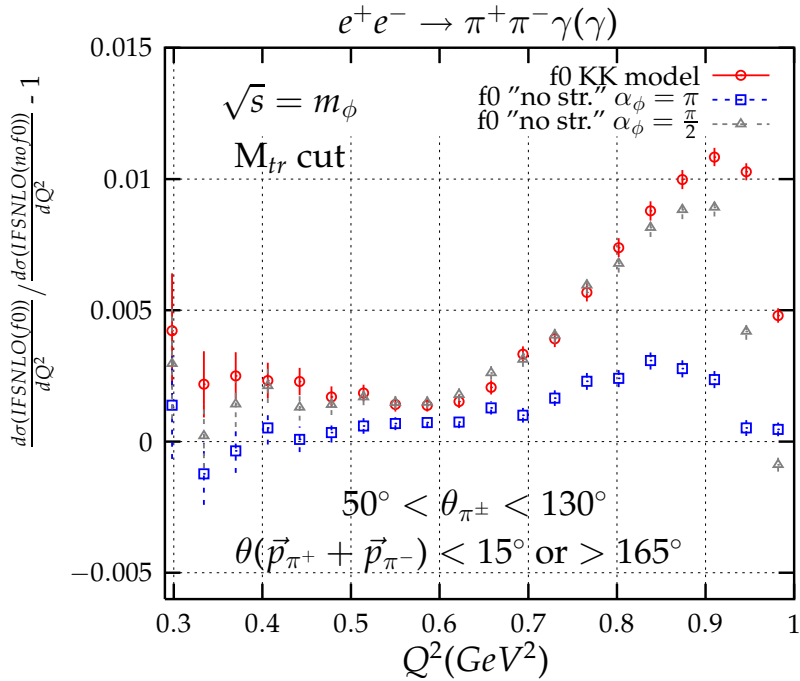}
\caption{The relative size of the direct radiative $\phi$ decay
 on the differential cross section for an event selection close
 to the one used by KLOE \cite{KLOE2}.
}
 \vspace{0.5 cm}
\label{f5}
\end{figure}
The answer can be found in Fig. \ref{f5}. With the models used 
in this paper one predicts  effects up to 1\% in the 
 $e^+e^-\to \pi^+\pi^-\gamma$ cross section 
 in the ``$KK$-model'' and the ``no- structure'' model, if
 $\alpha_\phi=$~$\pi$/2 is adopted, but significantly smaller values 
 for a different choice of the phase  $\alpha_\phi$
 (an event selection close
 to the one used by KLOE has been adopted). 
  All models are compatible with
  KLOE data on $\phi\to \pi^0\pi^0\gamma$ decay and thus
  the simultaneous analysis of $\phi\to \pi^0\pi^0\gamma$
 and $e^+e^-\to \pi^+\pi^-\gamma$ data is highly desirable.
 It will also allow for a full control of the FSR contributions
 to the $e^+e^-\to \pi^+\pi^-\gamma$ cross section, hopefully reducing
 the error
 from FSR in $\sigma(e^+e^-\to \pi^+\pi^-)$, as measured via the radiative
 return, to a negligible level.
\section{Summary and Conclusions}
\label{sec3}

The interplay between radiative corrections to pion pair production
 in electron-positron annihilation on one hand and $\pi\pi\gamma$
 production through radiative $\phi$-decays on the other hand
 has been investigated. The charge asymmetry is shown to be a unique
 signal for the interference between the amplitude for $\pi\pi$ production
 through the radiative return and $\pi\pi$ production from the radiative
 $\phi$-decay. This asymmetry can be studied as a function of the invariant
 mass of the $\pi\pi$-system and is particularly sensitive to the model
 for $\phi \to \pi\pi\gamma$. This has been demonstrated by implementing model
 amplitudes for $e^+e^-\to\phi \to \pi\pi\gamma$ into the PHOKHARA event
 generator and by studying the resulting distributions. Indeed different
 models, which lead to seemingly indistinguishable $\pi\pi$ mass distributions
 in $\phi$-decay, give rise to strikingly different charge asymmetries.
 We conclude with a discussion of the impact of radiative $\phi$-decays
 on the measurement of the pion form factor and demonstrate, that their
 effect can be kept well under control, if suitable kinematic
 regions are selected.

\section{Acknowledgments}

We would like to thank A.~Denig, W.~Kluge, D.~Leone and S.~M\"uller for 
discussions of the experimental aspects of our analysis. 
H.C. and A.G. are grateful for the support and the kind hospitality of
the Institut f{\"u}r Theoretische Teilchenphysik 
of the Universit\"at Karlsruhe.



\end{document}